\documentclass[10pt, letterpaper]{IEEEtran}
\usepackage{todonotes}
\usepackage[nocompress]{cite}
\usepackage{url}
\usepackage{enumitem}
\usepackage{epstopdf}
\PrependGraphicsExtensions{.tif, .tiff}
\usepackage{graphicx}
\usepackage{caption}
\usepackage{amsmath}
\usepackage{tabu}
\usepackage{multirow}
\usepackage{algorithm}
\usepackage{algpseudocode}
\usepackage{soul}
\usepackage[bookmarks=false]{hyperref}
\usepackage{footnote}
\usepackage{lipsum}
\usepackage{subcaption}
\usepackage{tikz}
\usetikzlibrary{positioning}
\soulregister\cite7
\soulregister\ref7
\soulregister\pageref7
\soulregister\secref7
\soulregister\tabref7
\newcommand\copyrighttext{%
  \footnotesize \copyright 2019 IEEE. Personal use of this material is permitted. Permission from IEEE must be obtained for all other uses, in any current or future media, including
reprinting/republishing this material for advertising or promotional purposes, creating new
collective works, for resale or redistribution to servers or lists, or reuse of any copyrighted
component of this work in other works.}
\newcommand\copyrightnotice{%
\begin{tikzpicture}[remember picture,overlay]
\node[anchor=south,yshift=10pt] at (current page.south) {\fbox{\parbox{\dimexpr\textwidth-\fboxsep-\fboxrule\relax}{\copyrighttext}}};
\end{tikzpicture}%
}

\newcommand{\scorep}{\mbox{Score-P}}

\newcommand{\figref}[1]{Figure~\ref{#1}}

\newcommand{\tabref}[1]{Table~\ref{#1}}

\newcommand{\secref}[1]{Section~\ref{#1}}

\begin{document}
\title{Modelling DVFS and UFS for Region-Based Energy Aware Tuning of HPC Applications}
\author{\IEEEauthorblockN{Mohak Chadha\IEEEauthorrefmark{1}, Michael Gerndt\IEEEauthorrefmark{2}}
\IEEEauthorblockA{\\Chair of Computer Architecture and Parallel Systems, Technische Universit{\"a}t M{\"u}nchen \\
Garching (near Munich), Germany} \\
Email: mohak.chadha@tum.de, gerndt@in.tum.de}

\maketitle

 \copyrightnotice

\begin{abstract}
Energy efficiency and energy conservation are one of the most crucial constraints for meeting the 20MW power envelope desired for exascale systems. Towards this, most of the research in this area has been focused on the utilization of user-controllable hardware switches such as per-core dynamic voltage frequency scaling (DVFS) and software controlled clock modulation at the application level. In this paper, we present a tuning plugin for the Periscope Tuning Framework which integrates fine-grained autotuning at the region level with DVFS and uncore frequency scaling (UFS). The tuning is based on a feed-forward neural network which is formulated using Performance Monitoring Counters (PMC) supported by x86 systems and trained using standardized benchmarks. Experiments on five standardized hybrid benchmarks show an energy improvement of 16.1\% on average when the applications are tuned according to our methodology as compared to 7.8\% for static tuning.             
% \todomc{Add results/feasibility of approach, results from comparison of static and dynamic switching.}
\end{abstract}
\begin{IEEEkeywords}
Energy-efficiency, autotuning, dynamic voltage and frequency scaling, uncore frequency scaling, dynamic tuning
\end{IEEEkeywords}

\IEEEpeerreviewmaketitle

\section{Introduction}
\label{sec:intro}
Modern HPC systems consists of millions of processor cores and offer petabytes of main memory. The Top500 list~\cite{Meuer2008} which is published twice every year ranks the fastest $500$ general purpose HPC systems based on their floating point performance on the LINPACK benchmark. The published data indicates that the performance of such systems has been steadily increasing along with increasing power consumption~\cite{top500},~\cite{green500}. The current fastest system delivers a performance of $122.30$ PFlop/s while consuming $8.81$MW of power. In order to achieve the $20$MW power envelope goal for an exascale system as published by DARPA~\cite{Bergman08exascalecomputing} we need $8.17$ fold increase in performance with only a $2.27$ fold increase in the power consumption.

With the growing constraints on the power budget, it is essential that HPC systems are energy efficient both in terms of hardware and software. Towards this, several power optimization techniques such as dynamic voltage and frequency scaling (DVFS), clock gating, clock modulation, ultra-low power states and power gating have been implemented in state-of-the-art processors by hardware vendors. DVFS is a common technique which enables the changing of the core frequency and voltage at runtime. It reduces the processor's frequency and voltage, resulting in reduced dynamic and static power consumption and thus leads to energy savings depending on the application characteristics. It is implemented using voltage regulators and dynamic clock sources by hardware vendors and is part of operating systems~\cite{dvfsondemand}. 

In earlier Intel processor architectures either the uncore frequency, i.e., frequency of uncore components (e.g. L3 cache), was fixed (Nehalem-EP and Westmere-EP) or the core and uncore parts of the processor shared a common frequency and voltage (Sandy Bridge-EP and Ivy Bridge-EP)~\cite{hswsurvey}. In new Intel processor architectures, i.e., Haswell onwards, a new feature called uncore frequency scaling (UFS) has been introduced. UFS supports separate core and uncore frequency domains and enables users to manipulate core and uncore frequencies independently. Changing the uncore frequency has a significant impact on memory bandwidth and cache-line transfer rates and can be reduced to save energy~\cite{hswsurvey},~\cite{Sourouri:2017:TFD:3126908.3126945}.
% \todomg{Should we write about system scenarios?}

%Introduce PTF
The Periscope Tuning Framework~\cite{periscope} is an online automatic tuning framework which supports performance analysis and performance tuning of HPC applications. It also provides a generic Tuning Plugin Interface which can be used for the development of plugins. In this paper, we present a tuning plugin which utilizes PTF's capability to manage search spaces and region-level optimizations to combine fine-grained autotuning with hardware switches, i.e., core frequency, uncore frequency and OpenMP~\cite{OpenMP} threads. The applications are tuned with node energy consumption as the fundamental tuning objective using a neural network~\cite{Haykin:2007:NNC:1213811} based energy model which is formulated using standardized PAPI counters~\cite{Mucci99papi:a} available on Intel systems. After the completion of performance analysis performed by PTF the tuning plugin generates a tuning model which contains a description of the best found hardware configurations, i.e., core frequency, uncore frequency and OpenMP threads for each scenario. A scenario consists of different regions which are grouped together if they have the same optimal configuration. This is commonly known as System-Scenario methodology~\cite{Gheorghita:2009:SDD:1455229.1455232} and is extensively used in embedded systems.
In order to evaluate our approach and the accuracy of our tuning model we use the generated tuning model as an input for the READEX Runtime Library (RRL) developed as part of the READEX project~\cite{readex}. RRL enables dynamic switching and dynamically adjusts the system configuration during application runtime according to the generated tuning model.
%Add info for overall results for energy efficiency after paper completion
%Add concept of rts in intro
% \todomc{Add concept of rts in intro}
Our key contributions are:
\begin{itemize}
    \item We implement and evaluate a model based tuning plugin for PTF with core frequency, uncore frequency and OpenMP threads as tuning parameters on HPC applications.
    \item We evaluate the accuracy of our model using sophisticated reference measurements conducted at high resolution.
    \item We practically demonstrate the viability of our methodology by comparing results for static and dynamically tuned applications and analyze performance-energy tradeoffs. 
    %think about 3rd
\end{itemize}

The rest of the paper is structured as follows. 
\secref{sec:background} provides a background on auto-tuners and describes the existing autotuning frameworks. In \secref{sec:tuning_plugin}, the tuning workflow and implementation of the plugin are outlined. \secref{sec:model} describes the modeling approach and the model used by the plugin. In \secref{sec:results}, the results of our approach are presented. Finally, \secref{sec:conclusion} concludes the paper and presents an outlook.

\section{Background and Related Work}
\label{sec:background}
A major challenge for several HPC applications is performance portability, i.e., achieving the same performance across different architectures. State-of-the-art autotuners aim to increase programmers productivity and ease of porting to new architectures by finding the best combination of code transformations and tuning parameter settings for a given system and tuning objective. The application can either be optimized according to a single criteria, i.e., single objective tuning or across a wide variety of criteria, i.e., multi objective tuning. 
%Added info about other tuning objectives, add more in Future work%
While this work focuses on energy consumption as the tuning objective, other objectives such as total cost of ownership (TCO), energy delay product (EDP) and energy delay product squared (ED2P) can also be used. Furthermore, the optimal configuration for tuning parameters can be determined for an entire application, i.e., static tuning or individually for each region by changing the tuning parameters dynamically at runtime, i.e., dynamic tuning. Since the search space created can be enormous, most autotuning frameworks utilize iterative techniques or analytical models to reduce the search space.

% Typical examples of tuning objectives include execution time, energy consumption, total cost of ownership (TCO) and energy delay product (EDP). 

% Typical examples of tuning parameters include compiler flags, user controllable switches and application specific parameters.

% Autotuning has been extensively utilized in special purpose libraries~\cite{CLINTWHALEY20013},~\cite{spiral} in order to optimize specific routines for a particular system. 
For optimizing parallel applications several autotuning frameworks which target different tuning parameters such as compiler flags~\cite{Triantafyllis:2003:COE:776261.776284},~\cite{autotune_comp} and application specific parameters~\cite{harmony} have been proposed. MATE~\cite{MORAJKO2007474} and ELASTIC~\cite{Martinez:2014:ELS:3184389.3184391} are two autotuning frameworks which support dynamic tuning of MPI based parallel applications. However, their approach is more coarse grained as compared to the one introduced in this paper using PTF~\cite{periscope}.

% Triantafyllis et al.~\cite{Triantafyllis:2003:COE:776261.776284} propose an iterative compilation technique called Optimization Space Exploration (OSE) for general purpose compilers. OSE utilizes heuristics, pruning and static performance estimation techniques to limit the different compiler performance optimization configurations present in the search space. Active Harmony~\cite{harmony} is a static autotuning framework which supports tuning of applications through automatic adaption of performance critical application-dependent parameters such as data distribution and algorithm-selection. 

% MATE~\cite{MORAJKO2007474} and ELASTIC~\cite{Martinez:2014:ELS:3184389.3184391} are two autotuning frameworks which support dynamic tuning of MPI based parallel applications. However, their approach is more coarse grained as compared to the one introduced in this paper using PTF~\cite{periscope}. 

While most autotuning frameworks focus on improving time-to-solution, some frameworks also support tuning of applications so as to improve energy efficiency~\cite{opentuner},~\cite{ptfdvfs}. Most of these techniques involve the use of DVFS. Guillen et al.~\cite{ptfdvfs} propose an automatic DVFS tuning plugin for PTF~\cite{periscope} which supports tuning of HPC applications at a region level with respect to power specialized tuning objectives. The plugin reduces the search by using analytical performance models and selects the optimal operating core frequency. Sourouri et al.~\cite{Sourouri:2017:TFD:3126908.3126945} propose an approach for dynamic tuning of HPC applications with respect to energy based tuning objectives. Similar, to this work their approach selects the optimal operating core, uncore frequency and the optimal number of OpenMP threads for tuning the application. However, the proposed approach involves an exhaustive search for finding the best configuration of tuning parameters, while in our approach we utilize an energy model to reduce the search space so as to select the optimal operating core and uncore frequency which significantly reduces the tuning time. Moreover in~\cite{Sourouri:2017:TFD:3126908.3126945}, each individual region needs to be identified and manually instrumented to incorporate DVFS and UFS switching, while in our approach significant region identification and application instrumentation are automatically achieved using \texttt{readex-dyn-detect}~\cite{readex} and \scorep{}~\cite{knupfer2012}. This allows our plugin to be used with any generic HPC application.

% However, the proposed approach involves extensive manual instrumentation of the application and an exhaustive search for finding the best configuration of tuning parameters. Contrary to the previous approach, we use an energy model to reduce the search space so as to select the optimal operating core and uncore frequency. Furthermore, our plugin can be used with a generic HPC application with minimal manual instrumentation. 

\section{Tuning Workflow}
\label{sec:tuning_plugin}
% \begin{figure*}[t]
% \centering
% \includegraphics[width=0.45\textwidth]{Images/tuning_plugin_workflow.png}
% \captionsetup{justification=centering}
% \caption{Overview of the Tuning Plugin Workflow}
% \label{fig:workflow}
% \end{figure*}
\begin{figure}[t]
\centering
\includegraphics[width=\columnwidth]{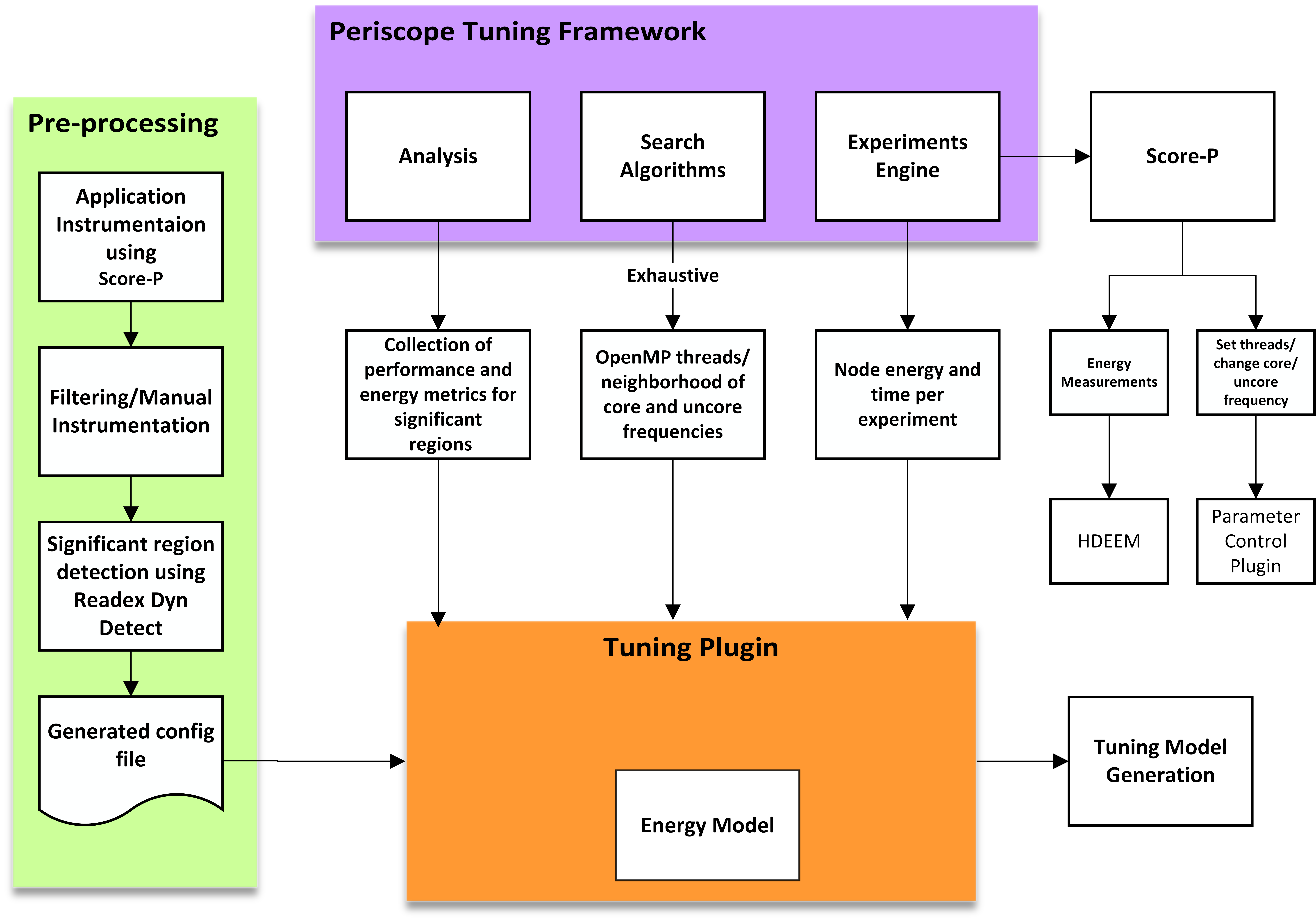}
\captionsetup{justification=centering}
\caption{Overview of the Tuning Plugin Workflow}
\label{fig:workflow}
\end{figure}
\figref{fig:workflow}. gives an overview of the different tuning steps involved in our tuning plugin workflow. The tuning of an application is primarily a four step process comprising of pre-processing, determination of optimal number of OpenMP threads, prediction of core and uncore frequencies for search-space reduction and tuning model generation. 
%Distinguishing each tuning phase
Steps one to four are offline and constitute the design time analysis (DTA) of the application. The production run which involves dynamic tuning using RRL~\cite{readex} is online and involves a lookup in the generated tuning model. This section describes each step in detail. 
% \todomc{Add Manual Instrumentation in Overflow}
\subsection{Pre-processing}
\label{pre_process}
The Periscope Tuning Framework~\cite{periscope} utilizes the measurement infrastructure for profiling and online analysis of HPC applications provided by \scorep{}~\cite{knupfer2012}. The first step involves compiler instrumentation of program functions, MPI and OpenMP regions using \scorep{}. This inserts measurement probes into the application's source code and can often lead to overheads during application execution. In order to reduce overhead, we filter the application using the tool \texttt{scorep-autofilter} developed as part of the READEX project~\cite{readex}. Filtering is a two step process and involves run-time and compile-time filtering. Executing the instrumented application with profiling enabled creates a call-tree application profile in the CUBE4 format~\cite{cube}. The application profile is then utilized during run-time filtering to generate a filter file which contains a list of finer granular regions below a certain threshold. The generated filter file is then used to suppress application instrumentation during compile-time filtering. An alternative method of reducing overhead is by using manual instrumentation. This requires analysis of the application profile generated by \scorep{} followed by manual annotation of the regions. 
% In this work, we use a threshold of $100$ms for removing finer granular regions.

Following this, we manually annotate the phase region of the application using specialized macros provided by \scorep{}. The phase region is a single entry, exit region which constitutes one iteration of the main program loop. The application is then executed to generate an application profile which is then used by the tool \texttt{readex-dyn-detect}~\cite{readex} to identify the different significant regions present in the application. A region qualifies as a significant region if it has a mean execution time of greater than $100$ms. Since energy measurement and application of core and uncore frequencies has a certain delay a threshold of $100$ms is selected to ensure that the right execution time influenced by setting the frequencies is measured. \texttt{Readex-dyn-detect} generates a configuration file containing a list of significant regions which is used as an input for the tuning plugin. 

\subsection{Tuning Step 1: Tuning OpenMP threads}
\label{tune_threads}
For tuning OpenMP and hybrid applications, the tuning plugin supports the number of OpenMP threads as a tuning parameter. The lower bound and step size for the tuning parameter can be specified in the configuration file generated at the end of the pre-processing step (see \secref{pre_process}). We use an exhaustive approach to determine the optimal number of OpenMP threads (see \figref{fig:workflow}). The tuning plugin creates scenarios depending upon the input in the configuration file, which are then executed and evaluated by the experiments engine. In order to dynamically change the number of OpenMP threads at runtime for each experiment, PTF utilizes the \scorep{} \texttt{OpenMPTP}\footnote{https://github.com/readex-eu/PCPs/tree/release/openmp\_plugin} Parameter Control Plugin (PCP)~\cite{readex}. The optimal number of OpenMP threads for each region are determined with energy consumption as the fundamental tuning objective. To obtain energy measurements we use the High Definition Energy Efficiency Monitoring (HDEEM)~\cite{Hackenberg:2014:HHD:2689710.2689711} infrastructure. HDEEM provides measurements at a higher spatial and temporal granularity with a sampling rate of (1 kSa/s) and leads to more accurate results. Energy measurement using HDEEM has a delay of $5$ ms on average. 

\subsection{Tuning Step 2: Tuning Core and Uncore frequencies}
\label{tune_freq}
Following the determination of optimal number of OpenMP threads, the tuning plugin requests the appropriate performance metrics for the phase region (see \secref{data:model_param}) in the analysis step as shown in \figref{fig:workflow}. These performance metrics are then used as an input for the energy model in the tuning plugin to predict energy consumption for different core and uncore frequencies. The combination of core and uncore frequency which leads to the minimum energy consumption is then used as the global core and uncore frequency. Global core and uncore frequency represent the optimal frequencies for the phase region and constitute the reduced search space. This highlights the main advantage of our modeling approach since we determine the optimal global core and uncore frequency in one tuning step.  Alternatively, the global operating core and uncore frequencies can also be determined by exhaustively searching through the parameter space. However, this significantly increases the tuning time. 
%We use the immediate neighboring frequencies of the global core and uncore frequency to exhaustively search for the optimal operating core and uncore frequencies for all significant regions. 
We use the immediate neighboring frequencies of the global core and uncore frequency to verfiy and select the operating core and uncore frequencies for all significant regions.
PTF utilizes the \scorep{} \texttt{cpu\_freq}\footnote{https://github.com/readex-eu/PCPs/tree/release/cpufreq\_plugin} and \texttt{uncore\_freq}\footnote{https://github.com/readex-eu/PCPs/tree/release/uncorefreq\_plugin} PCP plugins \cite{readex} to dynamically change core and uncore frequencies at runtime.

\subsection{Tuning Model Generation}
\label{tune_model}
% Will have to change this because we need to add concept of rts
% \todomc{Will have to change this because we need to add concept of rts}
After all experiments are completed and different system configuration parameters have been evaluated, the tuning plugin generates the tuning model. To avoid dynamic-switching overhead, regions which behave similar during execution or have the same configuration for different tuning parameters are grouped into scenarios by the plugin. This is done by using a classifier which maps each region onto a unqiue scenario based on its context. Each scenario lists the best found configuration of the hardware tuning parameters core frequency, uncore frequency and OpenMP threads. The generated tuning model is then used as an input for RRL~\cite{readex} (see \secref{stat_v_dyn}) for Runtime Application Tuning (RAT).

% The filtered application is then utilized for 
\section{Modelling Methodology}
\begin{figure}[t]
    \centering
    \begin{subfigure}{0.31\textwidth}
    \centering
        \includegraphics[width=\columnwidth,scale=1.5]{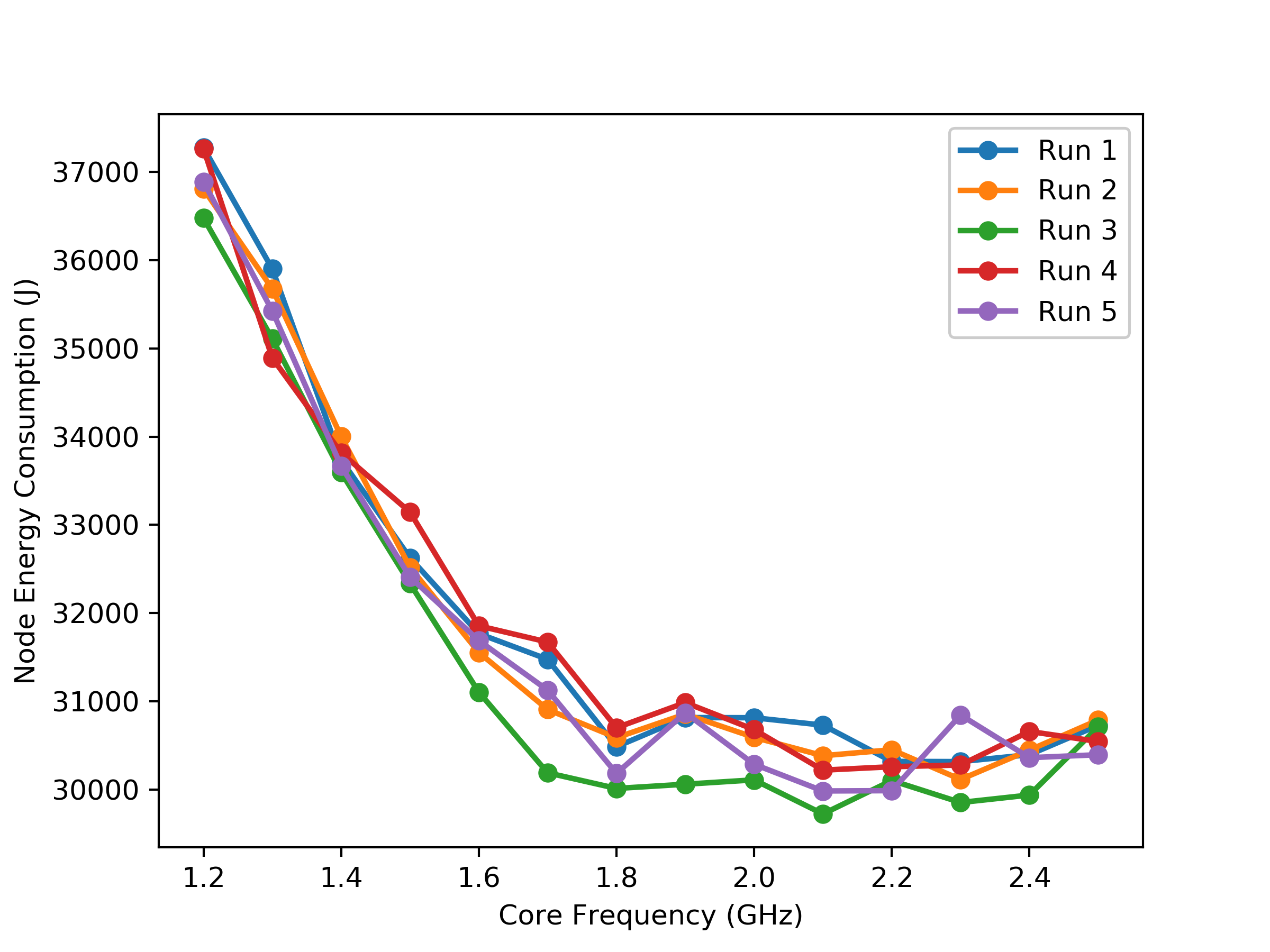}
        \caption{Node energy consumption with changing core frequency.}
        \label{fig_2}
    \end{subfigure}
    \begin{subfigure}{0.31\textwidth}
    \centering
        \includegraphics[width=\columnwidth,scale=1.5]{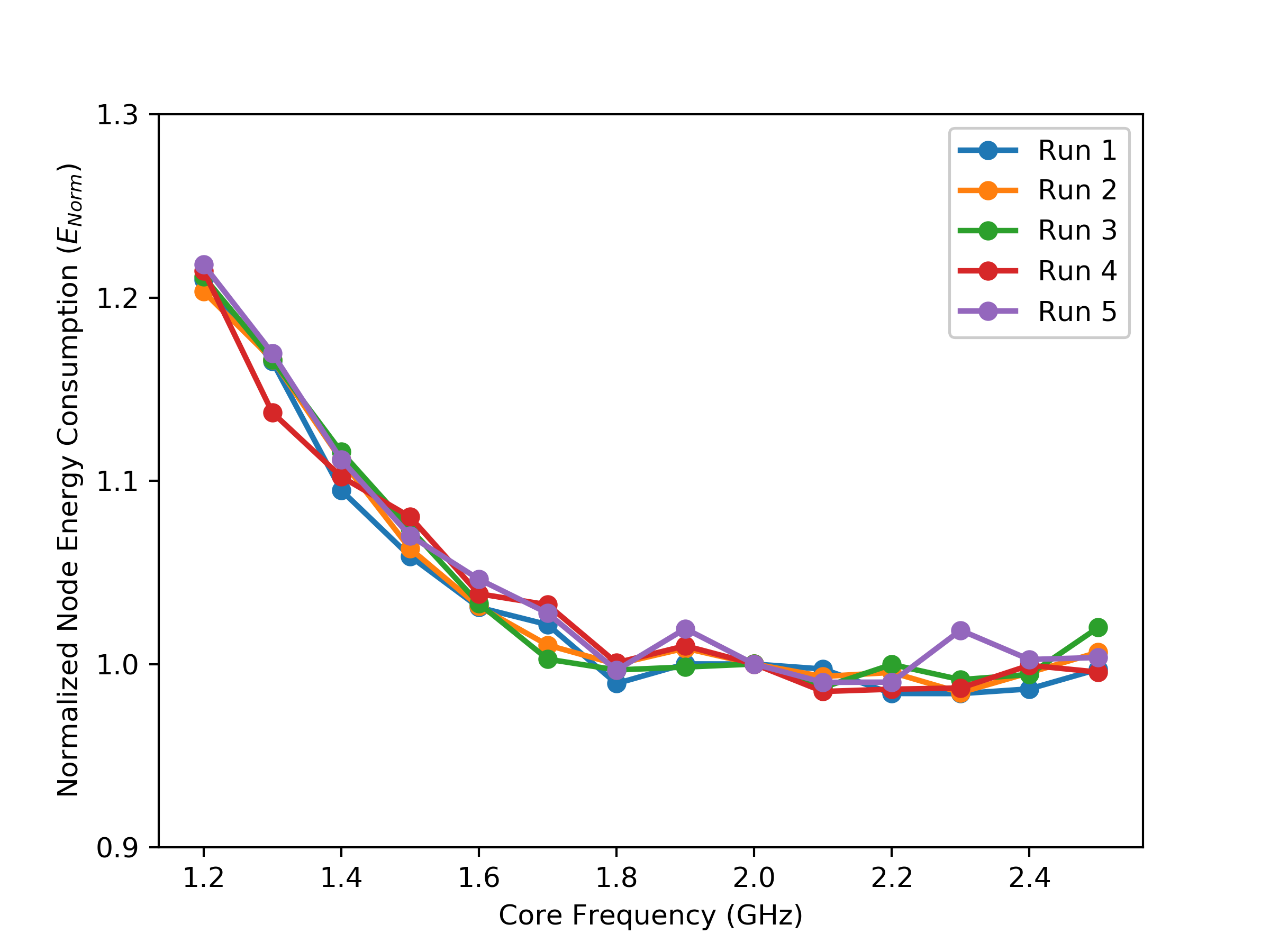}
        \caption{Normalized node energy consumption with changing core frequency.}
        \label{fig_3}    
    \end{subfigure}
    % \caption{Node energy and normalized node energy consumption for the application lulesh across different compute nodes}
    \caption{Comparison of node and normalized node energy consumption for the benchmark Lulesh across different compute nodes when uncore frequency is fixed.}
\end{figure}
\begin{figure}[t]
    \centering
    \begin{subfigure}{0.31\textwidth}
    \centering
        \includegraphics[width=\columnwidth,scale=1.5]{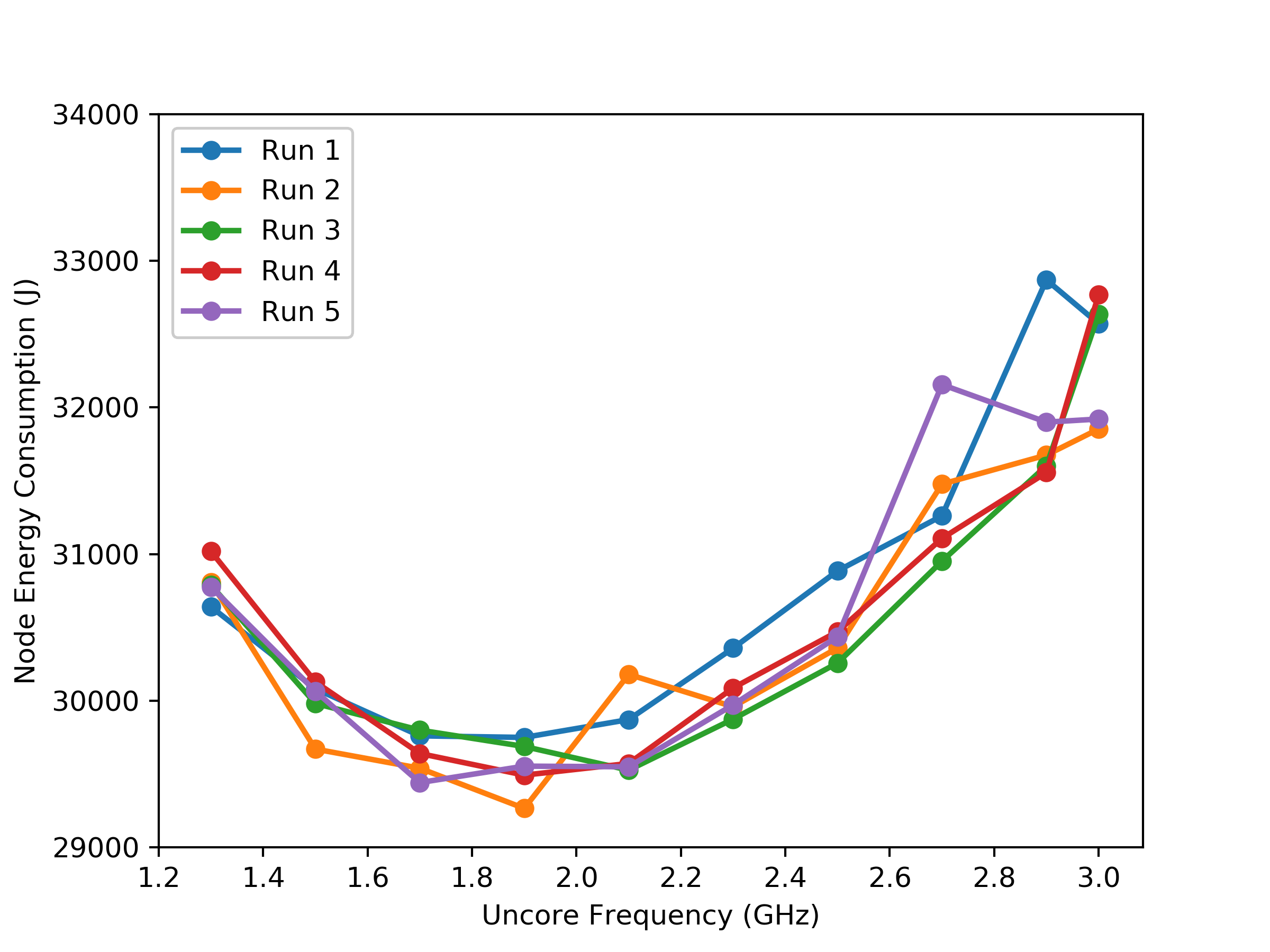}
        % \caption{MaxFlops}
        \caption{Node energy consumption with changing ucore frequency.}
        \label{fig_4}
    \end{subfigure}
    \begin{subfigure}{0.31\textwidth}
    \centering
        \includegraphics[width=\columnwidth,scale=1.5]{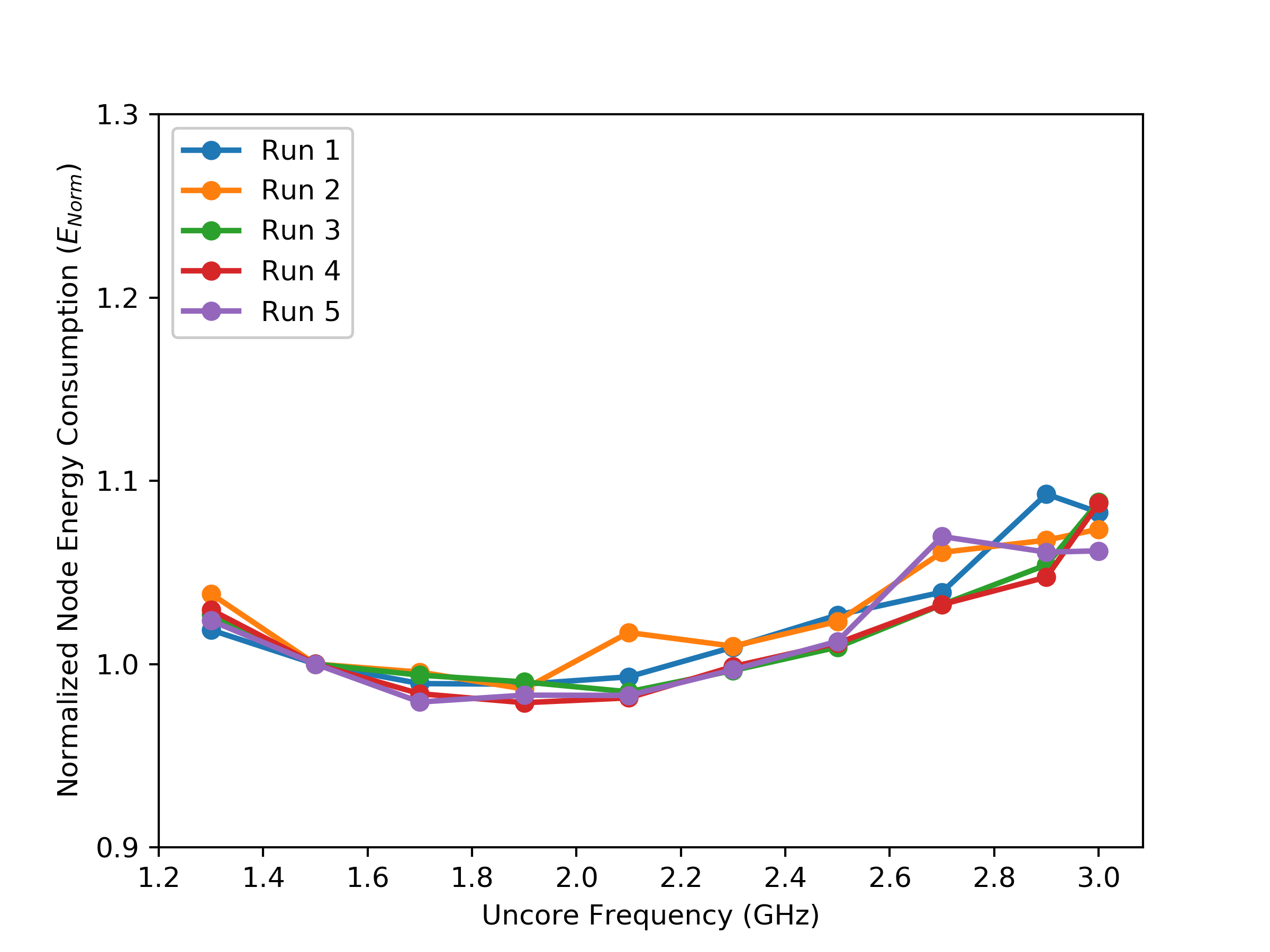}
        % \caption{FFT}
         \caption{Normalized node energy consumption with changing uncore frequency}
        \label{fig_5}
    \end{subfigure}
    \caption{Comparison of node and normalized node energy consumption for the benchmark Lulesh across different compute nodes when core frequency is fixed.}
\end{figure}
\label{sec:model}
Chadha et al.~\cite{power_stat_app} adapt and describe a statistically rigorous approach to formulate regression based power models for high performance x86 Intel systems, originally used for ARM processors~\cite{walker}. 
% The formulated models account for DVFS and use standardized PAPI~\cite{Mucci99papi:a} counters as independent variables.
The formulated models are based on DVFS frequencies and standardized PAPI~\cite{Mucci99papi:a} counters. In this paper, we extend their work to formulate energy models using a neural network architecture. This section describes our modelling approach in detail.

\subsection{Data Acquisition}
\label{data:acq}
% The first step involves collection of PAPI counter values along with energy information for different standardized benchmarks
% The first step involves co
Similar to~\cite{power_stat_app}, we use standardized PAPI counters for formulating our energy model. The first step involves the collection of these performance metrics and energy values for different HPC applications. The obtained PAPI counter values are then utilized to select an optimal subset of counters. We use the optimal PAPI counters as input for our energy model.

In order to obtain PAPI counter values, along with energy information for different standardized benchmarks we utilize the application tracing infrastructure provided by \scorep{}~\cite{knupfer2012}. We use \scorep{}'s built-in support for obtaining performance metrics to add PAPI data to the application trace. The energy values are added to the trace by using \texttt{scorep\_hdeem\_plugin}\footnote{https://github.com/score-p/scorep\_plugin\_hdeem} which implements the \scorep{} metric plugin interface. The performance metrics and energy values are recorded only at entry and exit of a region. The applications are then executed to generate the application trace in Open Trace Format 2 (OTF2)~\cite{Wagner2012} format.  The application trace consists of trace records sorted in a chronological order. To obtain energy values and performance metrics from application traces we implement a custom OTF2\footnote{https://github.com/kky-fury/OTF2-Parser} post-processing tool. Our tool reports energy values for the entire application run, while PAPI values are reported individually for instances of the phase region.
% while PAPI values are reported for one iteration of the phase region. 

Our experimental platform (see \secref{experimental_setup}) supports $56$ standardized PAPI counters along with $162$ native counters. Each native counter has many possible different configurations. We focus on the standardized PAPI counters to keep the amount of measurements needed feasible. For obtaining values of all $56$ PAPI counters, multiple runs of the same application are required due to hardware limitations on the simultaneous recording of multiple performance metrics. The energy and PAPI counter values are averaged across all runs. The core and uncore frequency values are fixed to $2.0$GHz and $1.5$ GHz for all measurements. 
% However, the measurements can be obtained at any particular value of core and uncore frequency, but typically the best values to choose are ones located in the middle of the available frequencies. 
Furthermore, we fix the OpenMP threads to $24$ for OpenMP and hybrid applications (see \tabref{table_2}).

\subsection{Model Parameter Selection}
\label{data:model_param}
A common pitfall of energy modelling is power variability among different compute nodes~\cite{beyonddvfs}. To examine this issue we consider two scenarios: 
\begin{enumerate}
    \item We vary the core frequency while keeping uncore frequency and OpenMP threads fixed at $1.5$GHz and $24$ respectively.
    \item We vary the uncore frequency  while keeping the core frequency and OpenMP threads fixed at $2.0$GHz and $24$ respectively.
\end{enumerate}

Scenarios $1$ and $2$ are shown for the benchmark Lulesh, executed with one MPI process and $24$ OpenMP threads on our experimental platform (see \secref{experimental_setup}) in \figref{fig_2} and \figref{fig_4} respectively. Each run in \figref{fig_2} and \figref{fig_4} indicates execution of the workload on a separate compute node multiple times with changing core and uncore frequency. 
% The energy profile of the application, when executed across different compute nodes, is fairly similar, as shown in \figref{fig_2} and \figref{fig_4}. However, the actual energy values depend upon the compute node where the application is being executed. 
The actual energy values of the application depend upon the compute node where the application is being executed as shown in \figref{fig_2} and \figref{fig_4}.
To account for the issue of power variability, we normalize the energy values of a particular run with energy value obtained at $2.0$GHz, $1.5$GHz core and uncore frequency respectively. 
This reduces the variability across runs on different compute nodes as shown in \figref{fig_3} and \figref{fig_5}. As a result of this, we train our model to predict normalized energy $E_{norm}$.

% We observe a maximum difference of $5$\% in \figref{fig_2} and $7$\% in \figref{fig_4}. To account for the issue of power variability, we normalize the obtained energy values with energy value at $2.0$ GHz $f_{clk}$ and $1.5$ GHz $f_{uclk}$. This reduces the variability to a maximum of $2$\% in Scenario $1$ and to $1.5$\% in Scenario $2$ as shown in \figref{fig_3} and \figref{fig_5} respectively. As a result of this, our model is trained to predict normalized energy $E_{norm}$. 

% The algorithm adapted by Chadha et al.~\cite{power_stat_app} for selecting PAPI counters to formulate power models is shown in~\algref{alg:select}
Chadha et al.~\cite{power_stat_app} describe an algorithm for selecting optimal PAPI counters for formulating regression based power models. The algorithm takes the entire set of standardized PAPI counters, obtained for a given set of workloads as input and returns an optimal set of counters. The counters are selected by formulating regression models between the PAPI counters and the dependent variable (power). We use the same algorithm for selecting optimal PAPI counters for our model with normalized node energy as the dependent variable. The authors also suggest the use of a heuristic criterion called Variance Inflation Factor (VIF) to quantify multicollinearity between the chosen PAPI counters. A large mean VIF value, usually greater than 10 indicates that the selected events are related to each other~\cite{kutner2004applied}. Collinearity between the selected events is a common pitfall of power and energy modelling~\cite{McCullough:2011:EEM:2002181.2002193}. To formulate stable models, it is essential that the selected counters are independent of each other so that the model can have maximum information regarding the workloads. 
\begin{table}[t]
\caption{Selected performance counters based on all workloads.}
	\centering
% 	\begin{tabu}{|c|c|c|c|}
    \begin{tabu}{|c|c|}
	\tabucline{-}
% 	\textbf{Counter} & \textbf{$R^{2}$} & \textbf{Adj.$R^{2}$} & \textbf{VIF} \\\tabucline{-}
% 	BR\_NTK & $0.517$ & $0.487$ & n/a  \\\tabucline{-}
% 	LD\_INS & $0.638$ & $0.591$ & $1.068$ \\\tabucline{-}
% 	L2\_ICR & $0.793$ & $0.748$ & $1.460$ \\ \tabucline{-}
% 	BR\_MSP & $0.863$ & $0.821$ & $1.587$ \\ \tabucline{-}
% 	RES\_STL & $0.921$ & $0.888$ & $2.405$ \\ \tabucline{-}
% 	SR\_INS & $0.967$ & $0.938$ & $2.941$ \\ \tabucline{-}
% 	L2\_DCR & $0.989$ & $0.972$ & $3.065$ \\ \tabucline{-}
    \textbf{Counter} & \textbf{mean VIF} \\\tabucline{-}
    BR\_NTK & n/a  \\\tabucline{-}
	LD\_INS & $1.068$ \\\tabucline{-}
	L2\_ICR & $1.460$ \\ \tabucline{-}
	BR\_MSP & $1.587$ \\ \tabucline{-}
	RES\_STL & $2.405$ \\ \tabucline{-}
	SR\_INS & $2.941$ \\ \tabucline{-}
	L2\_DCR & $3.065$ \\ \tabucline{-}
\end{tabu}
\label{tab:pmc_workloads}
\end{table}

To select the optimal counters for formulating our energy model, we run the standardized benchmarks (see \tabref{table_2}) on one compute node of our experimental platform (see \secref{experimental_setup}) with the system configuration described in \secref{data:acq}.
% at fixed core frequency of $2.0$GHz and uncore frequency of $1.5$GHz. The number of OpenMP threads are fixed to $24$ for OpenMP and Hybrid applications. 
%Explanation for CF 2.0 GHz and UCF 1.5 GHz
The optimal selected counters are shown in \tabref{tab:pmc_workloads}. The values of the selected counters depend only on the application characteristics and not on the frequencies. Hence any value of core and uncore frequency can be used for all measurements.  The obtained mean VIF for the selected counters is low indicating limited multicollinearity between the selected events. In \tabref{tab:pmc_workloads}, \texttt{BR\_NTK} describes the total number of conditional branch instructions not taken, \texttt{LD\_INS} describes the total number of load instructions, \texttt{L2\_ICR} describes the total number of L2 instruction cache reads, \texttt{BR\_MSP} describes the total number of mispredicted conditional branch instructions, \texttt{RES\_STL} describes the total number of cycles stalled on any resource, \texttt{SR\_INS} describes the total number of store instructions and \texttt{L2\_DCR} describes the total number of L2 data cache reads.

\subsection{Neural Network Architecture}
\label{data:neural_net_arch}
\tikzset{%
   neuron missing/.style={
    draw=none, 
    scale=4,
    text height=0.333cm,
    execute at begin node=\color{black}$\vdots$
  },
}
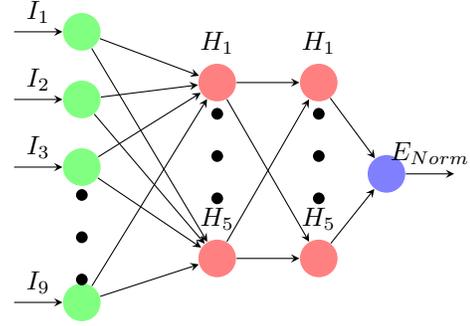
\begin{figure}[tb!]
\centering
\begin{tikzpicture}[x=0.9cm, y=0.9cm, >=stealth]

\foreach \m/\l [count=\y] in {1,2,3}
{
 \node [circle,fill=green!50,minimum size=0.5cm] (input-\m) at (0,2.5-\y) {};
}
\foreach \m/\l [count=\y] in {4}
{
 \node [circle,fill=green!50,minimum size=0.5cm ] (input-\m) at (0,-2.5) {};
}
 
 \node [neuron missing]  at (0,-1.5) {};

\foreach \m [count=\y] in {1}
  \node [circle,fill=red!50,minimum size=0.5cm ] (hidden-\m) at (2,0.75) {};
  
\foreach \m [count=\y] in {2}
  \node [circle,fill=red!50,minimum size=0.5cm ] (hidden-\m) at (2,-1.85) {};
  
 \node [neuron missing]  at (2,-0.3) {};

\foreach \m [count=\y] in {1}
  \node [circle,fill=red!50,minimum size=0.5cm ] (hidden_1-\m) at (3.5,0.75) {};
  
\foreach \m [count=\y] in {2}
  \node [circle,fill=red!50,minimum size=0.5cm ] (hidden_1-\m) at (3.5,-1.85) {};
  
 \node [neuron missing]  at (3.5,-0.3) {};

\foreach \m [count=\y] in {1}
  \node [circle,fill=blue!50,minimum size=0.5cm ] (output-\m) at (4.5,-0.60) {};
  
% \foreach \m [count=\y] in {2}
%   \node [circle,fill=blue!50,minimum size=1cm ] (output-\m) at (4,-0.5-\y) {};

%  \node [neuron missing]  at (4,-0.4) {};

\foreach \l [count=\i] in {1,2,3,9}
  \draw [<-] (input-\i) -- ++(-1,0)
    node [above, midway] {$I_{\l}$};

\foreach \l [count=\i] in {1,5}
  \node [above] at (hidden-\i.north) {$H_{\l}$};

\foreach \l [count=\i] in {1,5}
  \node [above] at (hidden_1-\i.north) {$H_{\l}$};

\foreach \l [count=\i] in {1}
  \draw [->] (output-\i) -- ++(1,0)
    node [above, midway] {$E_{Norm}$};

\foreach \i in {1,...,4}
  \foreach \j in {1,...,2}
    \draw [->] (input-\i) -- (hidden-\j);
    
\foreach \i in {1,...,2}
  \foreach \j in {1,...,2}
    \draw [->] (hidden-\i) -- (hidden_1-\j);
    
\foreach \i in {1,...,2}
  \foreach \j in {1}
    \draw [->] (hidden_1-\i) -- (output-\j);

%\foreach \l [count=\x from 0] in {Input, Hidden, Ouput}
 % \node [align=center, above] at (\x*2,2) {\l \\ layer};

\end{tikzpicture}
\caption{Used neural network architecture}
\label{fig:arch_nn}
\end{figure}
To formulate our energy model we use a 2-layer fully-connected neural network architecture as shown in \figref{fig:arch_nn}. The input layer consists of nine neurons, followed by two hidden layers consisting of five neurons and one neuron at the output layer. 
% The PAPI counters shown in \tabref{tab:pmc_workloads} are used as input to the model along with operating core and uncore frequencies. 
The PAPI counters shown in \tabref{tab:pmc_workloads} along with the operating core and uncore frequencies constitute the input features of our model. We standardize and center our input data by removing the mean and scaling to unit variance. This is primarily done to further reduce multicollinearity between the input features and to prevent one feature from dominating the model's objective function. The mean and scaling information is determined from the applications in our training set (see~\secref{sec:results}). PAPI counters are further normalized by dividing them with the execution time of one phase iteration as each application can have single or multiple phase iterations. 

% \begin{equation}
% \label{relu}
%     f(x) = max(0,x)
%     % \resizebox{.9\hsize}{!}f(x) = max(0,x)}
% \end{equation}
\begin{figure*}[t]
\centering
\includegraphics[width=\textwidth, scale=2.5]{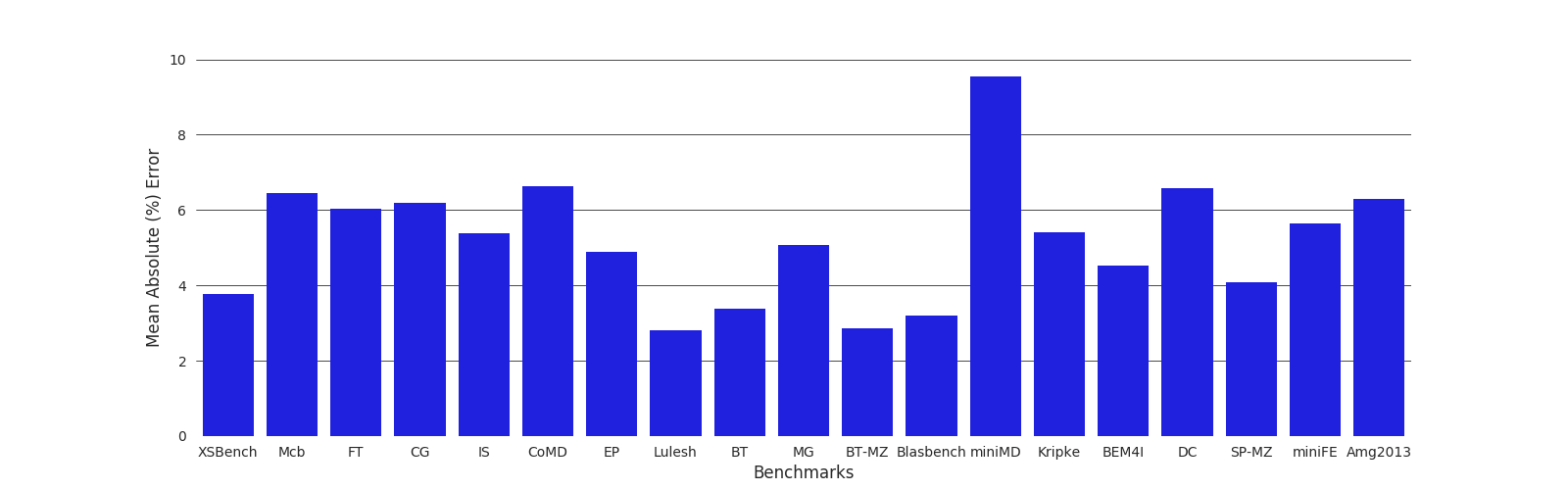}
\captionsetup{justification=centering}
\caption{Mean absolute (\%) error for 19 benchmarks across DVFS and UFS states when the network is trained using LOOCV technique.}
\label{fig:acc_results}
\end{figure*}

We use the Rectified Linear Units (ReLU)~\cite{Nair:2010:RLU:3104322.3104425} as activation functions in our neural network. This is because usage of ReLU has been proven to lead to faster convergence~\cite{Krizhevsky:2012:ICD:2999134.2999257} and they overcome the problem of vanishing gradients. The ReLU units are placed before the two hidden layers and before the output layer. 

We initialize the weights of the neurons in the network by randomly sampling from a zero mean, unit standard deviation Gaussian and multiplying with $\sqrt{(2.0/n)}$, as suggested by~\cite{DBLP:journals/corr/HeZR015}. Here $n$ represents number of neurons in the particular layer. This improves the rate of convergence and ensures that all neurons in the network initially have approximately the same output distribution. The biases of the neurons are initialized to zero. We use mean squared error as the objective function for training our model. The network predicts normalized energy $E_{norm}$ for a given set of PAPI counters, core and uncore frequency. In order to predict the global operating core and uncore frequency as described in \secref{tune_freq}, all combination of available frequencies are used as input to the network. The core and uncore frequency which leads to minimum energy consumption is then selected.

% \begin{equation}
% \label{relu}
%     W = np.random.randn(n) * sqrt(2.0/n)
%     % \resizebox{.9\hsize}{!}f(x) = max(0,x)}
% \end{equation}

\section{Experimental Results}
\label{sec:results}
In this section, we describe the system used for training and validating our energy model. We present results to demonstrate the accuracy of our models, compare static and dynamic tuning of applications and analyze energy performance trade-offs.

\subsection{System Description}
\label{experimental_setup}
For our experiments, we use the Bull cluster Taurus~\cite{Taurus} located at Technische Universit{\"a}t Dresden (TUD) in Germany. Taurus consists of six compute islands comprising of different Intel CPUs and Nvidia GPUs. Our experiments were performed on the \texttt{haswell} partition which consists of $1456$ compute nodes based on Intel Haswell-EP architecture. Each compute node has two sockets, comprising of two Intel Xeon E5-2680v3 processors with $12$ cores each and total $64$GB of main memory. Hyper-Threading and Turbo Boost are disabled on the system. The core frequency of each cpu core ranges from $1.2$GHz to $2.5$GHz, while the uncore frequency of the two sockets ranges from $1.3$GHz to $3.0$GHz. Each compute node is water cooled and is integrated with high resolution HDEEM~\cite{Hackenberg:2014:HHD:2689710.2689711} energy monitoring infrastructure. While running experiments, the energy measurements are obtained by using an FPGA integrated into the compute node which avoids perturbation and leads to high accuracy of energy measurements.   

\subsection{Neural Network Training}
\label{train:nn}
\begin{table}[t]
\caption{Benchmarks used for validation}
% 	\centering
% 	\begin{tabu}{|c|c|}
% 	\tabucline{-}
% 	\textbf{Suite} & \textbf{Benchmarks} \\\tabucline{-}
% 	NPB-3.3 & cg, dc, ep, ft, is, mg, bt-mz, sp-mz \\\tabucline{-}
% 	CORAL & amg2013, mcbenchmark, lulesh, miniFE, miniMD, XSBench, kripke \\\tabucline{-}
% 	Mantevo & CoMD, miniMD \\\tabucline{-}
% 	LLCBench & blasbench \\\tabucline{-}
% \end{tabu}
% \label{table_1}
\centering
\begin{tabular}{|c|c|} 
\hline
\textbf{Suite} & \textbf{Benchmarks} \\
\hline
NPB-3.3 & CG, DC, EP, FT, IS, MG, BT, BT-MZ, SP-MZ \\
\hline
% CORAL & \multirow{5}{6em}{amg2013, mcbenchmark, lulesh, miniFE, miniMD, XSBench, kripke} \\
\multirow{2}{4em}{CORAL} & Amg2013, Lulesh, miniFE, \\ & XSBench, Kripke, Mcbenchmark (Mcb) \\
\hline
Mantevo & CoMD, miniMD \\
\hline
LLCBench & Blasbench \\
\hline
Other & BEM4I \\
\hline
\end{tabular}
\label{table_2}
\end{table}

For training and validating our energy model experimentally, we use a wide variety of benchmarks from the NAS Parallel Benchmark (NPB) suite~\cite{Bailey:1991:NPB:125826.125925}, the CORAL benchmark suite~\cite{coral_suite}, the Mantevo benchmark suite~\cite{heroux2009improving}, LLCBench~\cite{llcbench} benchmark suite and a real world application BEM4I~\cite{MERTA2018106} which solves the Dirichlet boundary value problem for the 3D Helmholtz equation. The individual benchmarks are shown in \tabref{table_2}. The benchmarks selected from NPB except BT-MZ and SP-MZ along with miniFE (see \tabref{table_2}) are implemented using OpenMP. We use MPI only versions of the benchmarks Kripke and CoMD. All other benchmarks are implemented using both MPI and OpenMP, i.e., hybrid. 
% The benchmarks Kripke and CoMD are implemented using MPI, while all other benchmarks are implemented using both MPI and OpenMP, i.e., hybrid. 

To collect energy information at the different core and uncore frequencies, the applications are instrumented using \scorep{} and compiled using \texttt{gcc7.1}\footnote{Compiler flags: \texttt{-m64 -mavx2 -march=native -O3}} and \texttt{bullxmpi1.2.8.4}. We ran all the applications on one compute node of our experimental platform for all supported core frequencies and respectively uncore frequencies (see \secref{experimental_setup}) to generate \scorep{} OTF2~\cite{Wagner2012} traces. For OpenMP and hybrid applications we vary the number of OpenMP threads from $12$ to $24$ with a granularity of $4$. The generated traces are then post-processed to obtain energy information as described in \secref{data:acq}. The core and uncore frequencies are changed by using the low-level $x86\_adapt$~\cite{Schone:2014:IPA:2655100.2655109} library. We use $2.0$GHz, $1.5$GHz core and uncore frequency respectively for calibrating our energy model. The energy values are normalized by using the energy value of the particular application at the calibrating frequencies. Furthermore, the PAPI counter values obtained at the calibrating frequencies are used as input for the network.
% (see \secref{data:neural_net_arch}).

% The energy values are then normalized by using the energy value of the particular application at $2.0$GHz, $1.5$GHz core and uncore frequency respectively.

In order to evaluate the stability and performance of our model across unseen benchmarks we first train our network using the technique Leave-one-out cross-validation (LOOCV). In each step of LOOCV a single benchmark forms the testing set while the remaining benchmarks are used to train the network (see \tabref{table_2}). This step is repeated for all benchmarks. To train our network we use the stochastic optimization method ADAM~\cite{DBLP:journals/corr/KingmaB14}, which improves the rate of convergence. We use the default parameters of ADAM and a learning rate of $1e^{-3}$  for training our network. In each LOOCV step, the neural network is trained for five epochs, i.e. the neural network sees each training sample five times during forward and backward pass. Increasing the number of epochs greater than five leads to over-fitting and does not increase the accuracy of the model. \figref{fig:acc_results} shows the mean absolute (\%) error (MAPE) for all benchmarks (see \tabref{table_2}) across all DVFS and respectively UFS states. We obtain a maximum  MAPE value of $9.35$ for the benchmark miniMD and a minimum MAPE value of $2.81$ for the benchmark Lulesh. Our energy model achieves an average MAPE value of $5.20$ for all benchmarks as compared to $7.54$ achieved by the regression based power model, trained using 10-fold CV with random indexing in our previous work~\cite{power_stat_app}. A disadvantage of 10-fold CV is that some benchmarks might be repeated in both training and testing set. This indicates the stability and robustness of our energy model. Moreover, tuning for energy using regression would require two separate power and time models with core and uncore frequencies as independent variables, while this is accomplished by a simple 2-layer neural network as shown in this work.
% We achieve an average MAPE value of $5.20$ for all benchmarks which indicates the stability and robustness of our model. Our energy model achieves significantly better accuracy as 

Following this, we test our model for the hybrid benchmarks Lulesh, Amg2013, miniMD, BEM4I and Mcbenchmark and train using the rest. The neural network is trained for $10$ epochs with the same hyper-parameters used for LOOCV. In this scenario, we achieve a MAPE value of $7.80$ for the benchmarks in the test set. The weights and biases of the trained network are then used in the tuning plugin for region level tuning. 

\begin{figure}[t]
\centering
\includegraphics[width=0.35\textwidth,scale=1.5]{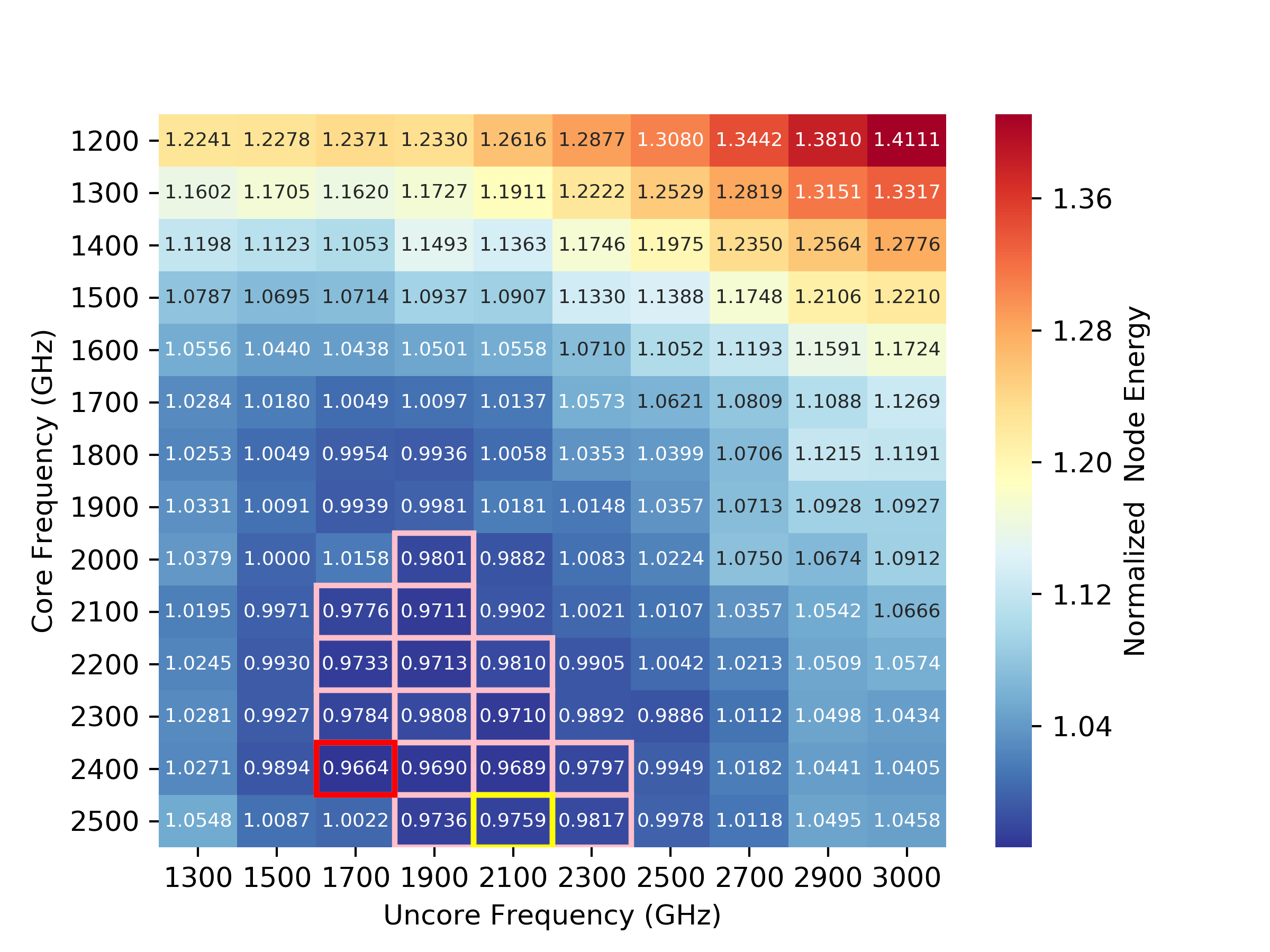}
\captionsetup{justification=centering}
\caption{Normalized node energy values for Lulesh for different core and uncore frequencies and $24$ OpenMP threads.}
\label{fig:lulesh_heatmap}
\end{figure}

\begin{table}[t]
\caption{Obtained optimal configuration for different significant regions of Lulesh}
\centering
\begin{tabular}{|>{\centering\arraybackslash}p{3.2cm}|>{\centering\arraybackslash}p{0.9cm}|>{\centering\arraybackslash}p{0.75cm}|>{\centering\arraybackslash}p{0.75cm}|} 
\hline
\textbf{Region}  & \textbf{OpenMP threads} & \textbf{CF \newline(GHz)} & \textbf{UCF (GHz)} \\
\hline
\footnotesize IntegrateStressForElems & $24$ & $2.50$ & $2.00$ \\ \hline
\scriptsize CalcFBHourglassForceForElems & $24$ & $2.50$ & $2.00$ \\ \hline
\footnotesize CalcKinematicsForElems &  $24$ & $2.40$ & $2.00$ \\ \hline
\footnotesize CalcQForElems & $24$ & $2.50$ & $2.00$ \\ \hline
\scriptsize ApplyMaterialPropertiesForElems  & $20$ & $2.40$ & $2.00$ \\ 
\hline
\end{tabular}
\label{table_5}
\end{table}

\subsection{Region-level tuning}
\label{region_level_tuning}
% \begin{figure}[t]
% \centering
% \includegraphics[width=0.40\textwidth,scale=1.5]{Images/heatmap_lulesh.png}
% \captionsetup{justification=centering}
% \caption{Normalized node energy values for the application Lulesh for different core and uncore frequencies and $24$ OpenMP threads.}
% \label{fig:lulesh_heatmap}
% \end{figure}

% \begin{table}[t]
% \caption{Obtained optimal configuration for different significant regions of Lulesh}
% \centering
% \begin{tabular}{|>{\centering\arraybackslash}p{3.2cm}|>{\centering\arraybackslash}p{0.9cm}|>{\centering\arraybackslash}p{0.75cm}|>{\centering\arraybackslash}p{0.75cm}|} 
% \hline
% \textbf{Region}  & \textbf{OpenMP threads} & \textbf{CF \newline(GHz)} & \textbf{UCF (GHz)} \\
% \hline
% \footnotesize IntegrateStressForElems & $24$ & $2.50$ & $2.00$ \\ \hline
% \scriptsize CalcFBHourglassForceForElems & $24$ & $2.50$ & $2.00$ \\ \hline
% \footnotesize CalcKinematicsForElems &  $24$ & $2.40$ & $2.00$ \\ \hline
% \footnotesize CalcQForElems & $24$ & $2.50$ & $2.00$ \\ \hline
% \scriptsize ApplyMaterialPropertiesForElems  & $20$ & $2.40$ & $2.00$ \\ 
% \hline
% \end{tabular}
% \label{table_5}
% \end{table}

The applications in the test set are instrumented with \scorep, pre-processed (see \secref{pre_process}) and then executed using PTF. We explicitly set the core and uncore frequency to the calibrating frequencies as described in \secref{train:nn} and run the applications with one MPI process and $24$ OpenMP threads. For the first tuning step (see \secref{tune_threads}) we use a lower bound and step size of $12$ and $4$ OpenMP threads respectively. 

\begin{figure}[t]
\centering
\includegraphics[width=0.35\textwidth,scale=1.5]{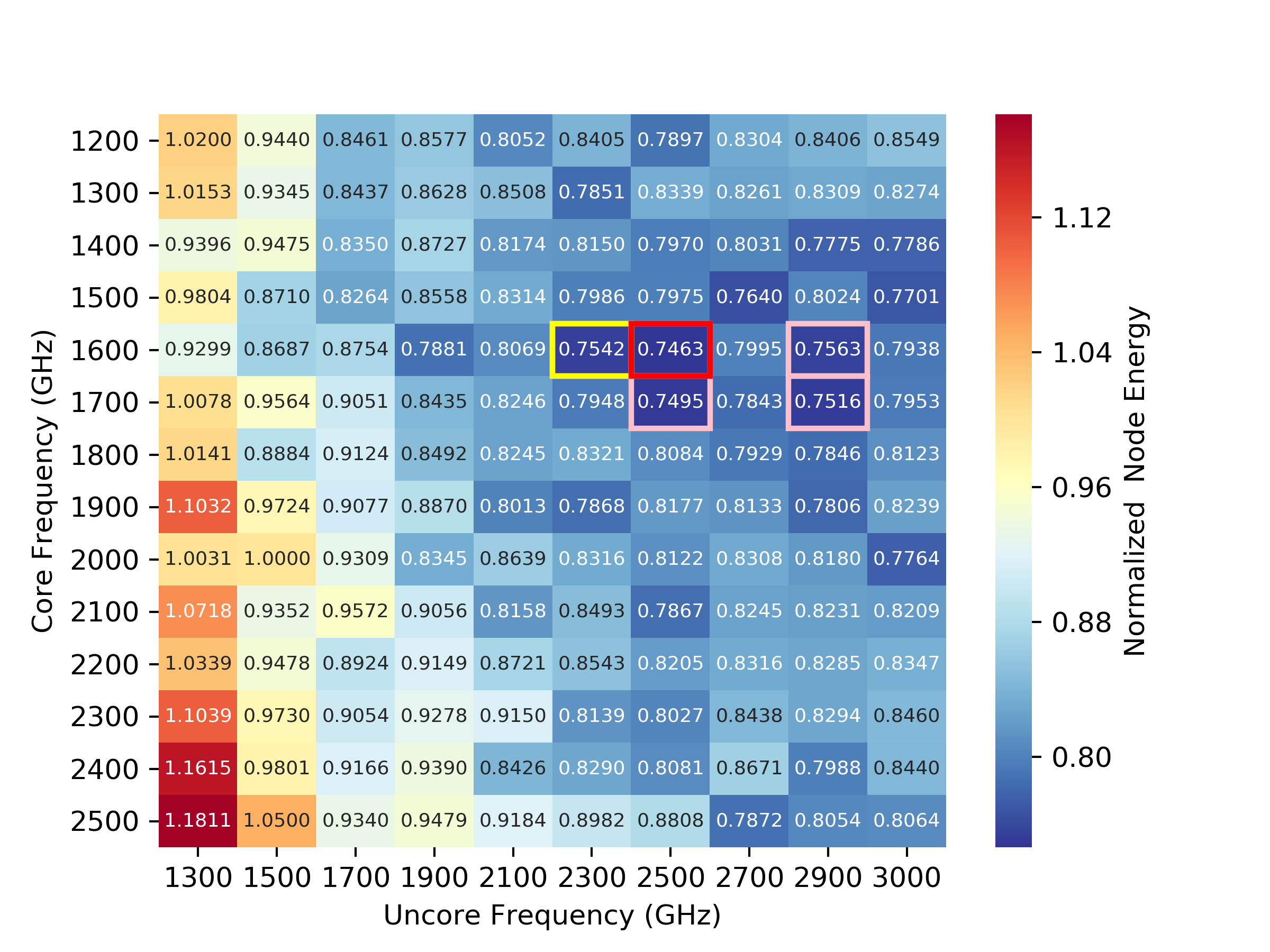}
\captionsetup{justification=centering}
\caption{Normalized node energy values for Mcbenchmark for different core and uncore frequencies and $20$ OpenMP threads.}
\label{fig:mcb_heatmap}
\end{figure}

\begin{table}[t]
\caption{Obtained optimal configuration for different significant regions of Mcbenchmark}
\centering
\begin{tabular}{|>{\centering\arraybackslash}p{3.2cm}|>{\centering\arraybackslash}p{0.9cm}|>{\centering\arraybackslash}p{0.75cm}|>{\centering\arraybackslash}p{0.75cm}|} 
\hline
\textbf{Region}  & \textbf{OpenMP threads} & \textbf{CF \newline(GHz)} & \textbf{UCF (GHz)} \\
\hline
setupDT & $24$ & $1.60$ & $2.30$ \\ \hline
advPhoton & $24$ & $1.60$ & $2.30$ \\ \hline
omp\_parallel:423 & $20$ & $1.60$ & $2.30$ \\ \hline
omp\_parallel:501 & $20$ & $1.70$ & $2.20$ \\ \hline
omp\_parallel:642 & $24$ & $1.60$ & $2.30$ \\ 
\hline
\end{tabular}
\label{table_6}
\end{table}

For the benchmark Lulesh the tuning plugin determines $24$ OpenMP threads as the optimum for the phase region. \figref{fig:lulesh_heatmap} shows the normalized node energy values for Lulesh at different core and uncore frequencies and $24$ OpenMP threads. With energy consumption as the fundamental tuning objective, \figref{fig:lulesh_heatmap} shows a trend towards higher core frequency and lower uncore frequency indicating that Lulesh is compute-bound. In \figref{fig:lulesh_heatmap} we highlight the best found configuration \texttt{2.4|1.7} GHz (core frequency (CF)$|$uncore frequency (UCF)) as red, the configuration selected by the tuning plugin \texttt{2.5|2.1} GHz (CF$|$UCF) as yellow and configurations withing $2$\% of the minimum as pink. Since, the normalized node energy values for different configuration of core and uncore frequencies (see \figref{fig:lulesh_heatmap}) are very close to the optimum, the actual energy values can vary across compute nodes and configurations which are not the most optimal can result in energy savings. The benchmark Lulesh consists of five significant regions which are determined using \texttt{readex-dyn-detect}~\cite{readex} (see \secref{pre_process}). The significant regions are different functions in Lulesh, names of which are shown in \tabref{table_5}. Following the selection of the optimal configuration for the phase region, the tuning plugin defines a reduced search space by using the frequencies in the immediate neighborhood of \texttt{2.5|2.1} GHz and generates scenarios. The OpenMP threads are fixed to the optimum obtained for the phase region. The scenarios are executed and evaluated by the experiments engine. The best found configuration for each significant region based on energy consumption is then selected. The best found configuration for each region is shown in \tabref{table_5}.

%Explain region tuning more detail 

% Following the selection of the optimal configuration for the phase region, the tuning plugin defines a reduced search space by using the frequencies in the immediate neighborhood of \texttt{2.5|2.1} GHz and generates scenarios. The OpenMP threads are fixed to the optimum for the phase region. The scenarios are executed and evaluated by the experiments engine. The best found configuration for each significant region based on energy consumption is then selected. The best found configuration for each region is shown in \tabref{table_5}.

% In the next tuning step, the plugin exhaustively determines the best configuration for each significant region based on energy consumption by searching in the immediate neighborhood of \texttt{2.5|2.1} GHz (CF$|$UCF). The best found configuration for each region is shown in \tabref{table_5}.

%Write about BEM4I, amg, miniMD

The tuning plugin determines the configuration $16$ OpenMP threads, \texttt{2.4|2.3} (CF$|$UCF) as the most optimal for the benchmark Amg2013 which consists of three significant regions. For the benchmark miniMD which also consists of three significant regions the configuration 24 OpenMP threads, \texttt{2.4|2.0} (CF$|$UCF) is found to be the most optimal. For the real world application BEM4I consisting of four significant regions, the configuration $24$ OpenMP threads, \texttt{2.4|2.4} (CF$|$UCF) is found to be the most optimal by the tuning plugin.

\begin{table}[t]
\caption{Obtained optimal static configuration}
\centering
\begin{tabular}{|>{\centering\arraybackslash}p{1.5cm}|>{\centering\arraybackslash}p{1.5cm}|>{\centering\arraybackslash}p{1.25cm}|>{\centering\arraybackslash}p{1.35cm}|} 
\hline
\textbf{Benchmark} & \textbf{OpenMP threads} & \textbf{CF \newline(GHz)} & \textbf{UCF (GHz)} \\
\hline
Lulesh & $24$ & $2.40$ & $1.70$ \\ \hline
Amg2013 & $16$ & $2.50$ & $2.30$ \\ \hline
miniMD & $24$ & $2.50$ & $1.50$ \\ \hline
BEM4I & $24$ & $2.30$ & $1.90$ \\ \hline
Mcbenchmark & $20$ & $1.60$ & $2.50$ \\ 
\hline
\end{tabular}
\label{table_3}
\end{table}

\begin{table*}[t]
\caption{Static and Dynamic Tuning Results}
\centering
\begin{tabular}{|>{\centering\arraybackslash}p{2cm}|>{\centering\arraybackslash}p{3.5cm}|>{\centering\arraybackslash}p{5cm}|>{\centering\arraybackslash}p{4cm}|}
\hline
% \multicolumn{1}{|>{\centering\arraybackslash}p{2cm}|}{\textbf{Benchmark}} 
    \multirow{3}{*}{\textbf{Benchmark}}
    & \multicolumn{1}{>{\centering\arraybackslash}p{3.5cm}|}{\textbf{Static tuning savings}} 
    & \multicolumn{2}{c|}{\textbf{Dynamic tuning savings}} \\
    \cline{2-4}
& job energy/CPU energy/time & job energy/CPU energy/time/performance reduction config setting & overhead DVFS/UFS/Score-P \\
\hline

Lulesh & $1.14$\%/$2.60$\%/$0.97$\% & $5.48$\%/$10.30$\%/$-7.70$\%/$-5.46$\% & $-2.24$\% \\ \hline 
Amg2013 & $4.89$\%/$12.63$\%/$-6.80$\% & $5.42$\%/$16.67$\%/$-11.2$\%/$-8.96$\% & $-2.24$\%\\ \hline
miniMD & $4.10$\%/$8.63$\%/$0.41$\% & $10.3$\%/$21.95$\%/$-4.00$\%/$-2.29$\% & $-1.71$\% \\ \hline
BEM4I & $2.64$\%/$4.61$\%/$0.70$\% & $8.26$\%/$12.43$\%/$-4.25$\%/$-2.98$\%& $-1.27$\% \\ \hline
Mcbenchmark & $6.00$\%/$10.50$\%/$-6.50$\% & $8.20$\%/$18.76$\%/$-14.50$\%/$-10.10$\%&$-4.40$\% \\ \hline
\end{tabular}
\label{table_4}
\end{table*}

While the above discussed benchmarks are compute bound, Mcbenchmark is predominantly memory bound. \figref{fig:mcb_heatmap} shows a trend towards higher uncore frequency and lower core frequency indicating the need for higher memory bandwidth. The tuning plugin determines $20$ OpenMP threads, \texttt{1.6|2.3} GHz (CF$|$UCF) as the optimal configuration of the phase region as compared to the optimum at \texttt{1.6|2.5} GHz (CF$|$UCF) as shown in \figref{fig:mcb_heatmap}. Mcbenchmark consists of five significant regions, two functions and three OpenMP parallel constructs. The optimal configuration for each significant region is shown in \tabref{table_6}. At the end of the PTF run the tuning plugin generates the tuning model by grouping regions with similar configuration into scenarios as described in \secref{tune_model}. 

In order to quantify the tuning time in our approach as compared to the one introduced in~\cite{Sourouri:2017:TFD:3126908.3126945}, consider the workload Mcbenchmark with $n$ regions. Suppose that one run of the benchmark takes $t$ sec and the search space for finding the optimal configuration, i.e.,  OMP$|$CF$|$UCF is $k$ x $l$ x $m$. Since the approach introduced in~\cite{Sourouri:2017:TFD:3126908.3126945} does not consider significant regions and uses an exhaustive search policy, the tuning time would be $n$ x $k$ x $l$ x $m$ x $t$. However, since we tune all significant regions in a single application run and use an energy model in our approach, the tuning time is significantly reduced to ($k$ + $1$ + $9$) x $t$ as discussed in \secref{sec:tuning_plugin}. Moreover, in applications with progressive loops such as Lulesh each phase iteration can be exploited and the entire application run is not required. In that case, the tuning time would be ($k$ + $1$ + $9$).

\subsection{Comparing Static and Dynamic Tuning}
%Complete tomorrow
\label{stat_v_dyn}
 To compare static and dynamic tuning, we consider three parameters job energy, CPU energy and time. To measure job energy and time, we use the SLURM tool \texttt{sacct} which allows users to query post-mortem job data for any previously executed jobs or job steps. The energy and time values can be obtained by using the \texttt{--format} parameter. For measuring CPU energy we utilize a lightweight runtime tool called \texttt{measure-rapl} which uses the $x86\_adapt$~\cite{Schone:2014:IPA:2655100.2655109} library to measure the CPU energy via Intel's RAPL interface. 

\tabref{table_3} shows the optimal static configuration found for each benchmark. These values are obtained by running the benchmarks at different OpenMP threads, core frequencies and uncore frequencies. The configuration which results in minimum energy consumption is then selected. The best found static configuration is equivalent to the best configuration found for the phase region. The default operating core and uncore frequency for any job running on our experimental platform (see \secref{experimental_setup}) is \texttt{2.5|3.0} GHz (CF$|$UCF). In order to compute static savings for a particular benchmark, the benchmark is first executed with a default configuration of $24$ OpenMP threads and $2.5|3.0$ GHz (CF$|$UCF) on a compute node. Following this, we manually set the best obtained static configuration (see \tabref{table_3}) and execute the benchmark on the same compute node. Savings in terms of job energy, CPU energy and time are then computed relative to the values for the default configuration. 

Following the tuning model generation in the previous step for the five instrumented benchmarks, we use the tuning model as an input for the RRL\footnote{https://github.com/readex-eu/readex-rrl}~\cite{readex} library. This is done by using the environment variable \texttt{SCOREP\_RRL\_TMM\_PATH}. RRL enables runtime application tuning and uses the \scorep{} PCP plugins to dynamically change the system configuration at runtime. The configuration applied is extracted from the different scenarios in the generated tuning model (see \secref{tune_model}). The job energy, CPU energy and time values at the end of the RRL run are then used to compute dynamic savings relative to the parameter values for the default configuration. The values for both static and dynamic savings are averaged over five runs.

\tabref{table_4} shows the static and dynamic tuning savings for the five benchmarks. For static tuning we achieve average savings of $3.5$\%, $7.8$\% in terms of job and CPU energy respectively. On the other hand, when the benchmarks are dynamically tuned using our methodology, average energy improvement of $7.53$\%, $16.1$\% for job and CPU energy is observed. The increase in energy savings can be attributed to the selection of only certain regions above a threshold as significant (see \secref{pre_process}) and the usage of scenarios in the tuning model. Maximum energy improvement of $10.3$\%, $21.95$\% in terms for job and CPU energy is observed for the dynamically tuned benchmark miniMD, while minimum is observed for the benchmark Lulesh with static tuning (see \tabref{table_4}). Static tuning leads to a slight improvement in performance for the benchmarks Lulesh, miniMD and BEM4I, while the performance is decreased for the benchmarks Amg2013 and Mcbenchmark. For Amg2013 the decrease in performance can be attributed to the optimal static configuration of $16$ OpenMP threads as compared to $24$ for the default configuration. In case of Mcbenchmark the increase in execution time is primarily due to the optimal low operating core frequency value of $1.6$ GHz (see \tabref{table_3}) in comparison to $2.5$ GHz for the default configuration. 

\subsection{Analyzing Overhead}
\label{sec:analysis}
Although dynamic tuning is more energy efficient, it leads to a decrease in performance as shown in \tabref{table_4}. The decrease in performance is because of three reasons. First, reduction in performance due to configuration setting. Second, overhead due to dynamic switching. Third, instrumentation overhead due to \scorep{}. To quantify the performance reduction due to the configuration setting, we measure the relative execution time of each reach region w.r.t the default configuration for each benchmark. The values found are shown in \tabref{table_4}. To change the core and uncore frequencies the \scorep{}  PCP plugins utilize the $x86\_adapt$~\cite{Schone:2014:IPA:2655100.2655109} library. The transition latency for changing frequency of one individual core on our experimental platform is $21\mu s$, while changing the 
operating uncore frequency for each socket has a transition latency of $20\mu s$. Without considering scenarios, the DVFS/UFS overhead for a particular benchmark can be computed by multiplying the number of iterations with the number of significant regions and the transition latency. Although runtime and compile-time filtering along with manual instrumentation (see \secref{pre_process}) reduce \scorep{} overhead to a large extent, it is not completely removed due to instrumentation of OpenMP and MPI routines by \scorep{}. The combined DVFS/UFS/\scorep{} overhead in our approach ranges from $20$-$30.34$\% for the five benchmarks (see \tabref{table_4}) as compared to $10$-$50$\% in~\cite{Sourouri:2017:TFD:3126908.3126945}.

\section{Conclusion \& Future Work}
\label{sec:conclusion}
Energy-efficiency of current as well as future HPC systems remains a key challenge due to their increasing number of components and complexity. As a result, several techniques which utilize dynamic voltage and frequency scaling, software clock modulation and power-capping to tune and reduce the energy consumption of applications have been developed. However, most of these techniques either require extensive manual instrumentation or tune the applications statically. In this paper, we have developed a tuning plugin for the Periscope Tuning Framework which utilizes user-controllable hardware switches, i.e., OpenMP threads, core frequency and uncore frequency to automatically tune HPC applications at a region level.   

The tuning plugin consists of two tuning steps. In the first tuning step, the optimal number of OpenMP threads for each significant region are exhaustively determined. Following this, the tuning plugin utilizes a neural-network based energy model to predict the optimal operating core and uncore frequency in one tuning step. 
% This reduces the search space and allows faster tuning of applications in comparison to exhaustively exploring the entire search space. 
The tuning plugin then uses the immediate neighbors of the obtained core and uncore frequency to verify and select the optimal configuration for each significant region. The search time for finding the optimal configuration is significantly reduced by using an energy model and evaluating for each individual phase iteration. In comparison, it is not required to run the entire applications for experiments in the exhaustive search space. 

The energy model is formulated using standardized PAPI counters available on Intel systems and accounts for two common pitfalls of energy modelling on HPC systems, power variability and multicollinearity between the selected events. Furthermore, we demonstrate the accuracy and stability of our model across $19$ standardized benchmarks by using the technique Leave-one-out-cross-validation, achieving an average MAPE value of $5.20$ for all DVFS and respectively UFS states.

We present results for region based tuning of five hybrid applications. Moreover, we demonstrate the viability of our approach by comparing static and dynamically tuned versions of the applications. For dynamic tuning we utilize the READEX Runtime Library (RRL) which dynamically changes the system configuration at runtime. Our experiments show an energy improvement of $7.53$\%, $16.1$\% in terms of job and CPU energy for dynamically tuned applications as compared to $3.5$\%, $7.8$\% for static tuning. 

In the future we want to investigate the application of the model based approach to individual significant regions. By that regions with a very different best configuration could be identified, e.g., IO regions. Furthermore, we would like to add support for other energy based tuning objectives such as EDP, ED2P.
% The future work of our project will focus on extending our work to runtime situations (rts) and using an energy model to individually predict the optimal configuration for each rts. The context element of a significant region is referred to an rts. 

% The energy model is formulated using standardized PAPI counters available on Intel systems,  
% The neural network is formulated using standardized PAPI counters available on Intel systems and takes $9$ inputs. We account for two common pitfalls of energy modelling on HPC systems power variability and multicollinearity between the selected events. Furthermore, we demonstrate the accuracy and stability of our model 

\section{Acknowledgment}
\label{sec:ack}
% The research leading to these results has received funding from the European Union’s Horizon 2020 Programme under grant agreement number 671657. We thank the Centre for Information Services and High Performance Computing (ZIH) at TU Dresden for providing HPC resources that contributed to our research.
The research leading to these results was partially funded by the European Union’s Horizon 2020 Programme under grant agreement number 671657 and Deutsche Forschungsgemeinschaft 
(DFG, German Research Foundation)-Projektnummer 146371743-TRR 89: Invasive Computing.
% the German Research Foundation (DFG) as part of the Transregional
% Collaborative Research Centre “Invasive Computing” (SFB/TR 89). 
We thank the Centre for Information Services and High Performance Computing (ZIH) at TU Dresden for providing HPC resources that contributed to our research. 

% These power saving techniques are primarily used for implementing the four different power saving states, i.e, system sleeping states (S-states), processor power states (C-states), processor performance states (P-states) and throttling states (T-states) supported by Intel processors as defined by the Advanced Configuration and Power Interface (ACPI) Specification \cite{acpi}.

\bibliographystyle{IEEEtran}
\bibliography{parallelpgm}

% \listoftodos

\end{document}